\documentclass[pss]{wiley2sp} 
\usepackage{amsmath}
\usepackage{cite}
\usepackage{color}
\usepackage{graphicx}
\usepackage{bm}              
\usepackage{w-greek}         

\begin{document}

\title{Electronic comparison of InAs wurtzite and zincblende phases using nanowire transistors}
\titlerunning{Wurtzite and Zincblende InAs nanowire transistors}

\author{%
  A.R.~Ullah\textsuperscript{\textsf{\bfseries 1}},
  H.J.~Joyce\textsuperscript{\textsf{\bfseries 2}},
  A.M.~Burke\textsuperscript{\textsf{\bfseries 1}},
  H.H.~Tan\textsuperscript{\textsf{\bfseries 2}},
  C.~Jagadish\textsuperscript{\textsf{\bfseries 2}},
  A.P.~Micolich\textsuperscript{\Ast,\textsf{\bfseries 1}}}

\authorrunning{Ullah {\it et al.}}

\mail{e-mail
  \textsf{adam.micolich@nanoelectronics.physics.unsw.edu.au}, Phone: +61 2 9385 6132}

\institute{%
  \textsuperscript{1}\,School of Physics, University of New South Wales, Sydney NSW 2052, Australia\\
  \textsuperscript{2}\,Department of Electronic Materials Engineering, Research School of Physics and Engineering, The Australian National University, Canberra ACT 0200, Australia\\}

\received{XXXX, revised XXXX, accepted XXXX} 
\published{XXXX} 


\abstract{\abstcol{We compare the electronic characteristics of
nanowire field-effect transistors made using single pure wurtzite
and pure zincblende InAs nanowires with nominally identical
diameter. We compare the transfer characterist-}{ics and
field-effect mobility versus temperature for these devices to better
understand how differences in InAs phase govern the electronic
properties of nanowire transistors.}}

\maketitle

\section{Introduction}

The small volume and surface area of semiconductor nanowires enables
high-quality interfaces that are difficult/impossible in bulk
structures~\cite{SamuelsonMT03}. The InAs/InP~\cite{BjorkAPL02} or
Si/Ge~\cite{LauhonNat02} heterointerface is the classic example;
however, crystal phase homointerfaces in a single semiconductor are
also possible. In InAs nanowires these are the zincblende (ZB) phase
observed in bulk III-Vs, and the wurtzite (WZ)
phase~\cite{HirumaJAP95}. This phase mixing is usually random and
considerable effort has been invested into phase-engineered
nanowires~\cite{AlgraNat08, CaroffNatNano09, DickNL10,
CaroffIEEE-JQSTE11}. One key motivation is using the different band
alignments~\cite{MurayamaPRB94, ZanolliPRB07, DePRB10} to make
devices such as quantum dots~\cite{SpirkoskaPRB09, AkopianNL10}.
Another is obtaining phase-pure WZ or ZB nanowires to prevent
phase-interface scattering from degrading electrical
performance~\cite{IkonicPRB93, ThelanderNL11}.

Studies of how nanowire crystal phase affects electrical properties
are at an early stage. Dayeh {\it et al.} reported characterization
at temperature $T = 300$~K of InAs nanowire field-effect transistors
(NWFETs) made using pure ZB nanowires and WZ nanowires with small ZB
segments interspersed axially (approx. $3.5$~nm ZB per $28.5$~nm
WZ)~\cite{DayehAFM09}. While the mobility $\mu$ was comparable, the
ZB NWFETs had higher off-current giving a poor on-off ratio
$I_{\textrm{on}}/I_{\textrm{off}} \sim 2$ compared to $10^{4}$ for
the WZ NWFETs. The difference was attributed to spontaneous
polarization charges at the WZ/ZB interfaces. Schroer {\it et al.}
studied single nanowires with low ($<1~\mu$m$^{-1}$) and high
($>1$~nm$^{-1}$) defect density segments~\cite{SchroerNL10}. The
mobility at $T = 4.2$~K was $\sim 4\times$ larger for the
defect-free segments; while not a direct comparison between WZ and
ZB, it shows the potential mobility gains achievable with phase-pure
nanowires. Sladek {\it et al.} studied InAs nanowire conductivity
for three growth methods: one producing WZ nanowires with high
stacking fault density, the other two producing ZB
nanowires~\cite{SladekPhysStatSolC12}. The measurements showed that
doping, intentional or otherwise, dominates over crystal structure
in determining conductivity.

Joyce {\it et al.} recently developed a growth method for phase-pure
WZ and ZB nanowires (NWs) without limiting diameter choice or
requiring dopant addition; control was exerted using temperature and
V/III ratio~\cite{JoyceNL10}. Conditions for obtaining phase-pure
InAs nanowires are now well established~\cite{DickNL10,
DickJVSTB11}. Comparative studies of how phase affects thermal
conductivity~\cite{ZhouPRB11} and optical
properties~\cite{SunNanotech10, WilhelmNanoscale12} have been
reported, as has an electrical study involving phase-pure InAs
NWs~\cite{ThelanderNL11}. Thelander {\it et al.} studied resistivity
$\rho$ versus phase fraction from WZ NWs with low stacking fault
density to ZB NWs with periodic twinning structure. For nanowires
grown by metalorganic vapor phase epitaxy (MOVPE), phase control was
exerted by changing the diameter from $40$ (WZ) to $120$~nm (ZB) at
fixed $T$ and V/III ratio. WZ NWs were also grown by molecular beam
epitaxy (MBE) to account for background doping and diameter effects.
The addition of even small amounts of WZ phase to ZB NWs have a
significant effect, increasing $\rho$ by two orders of magnitude.
Further, pure ZB NWs had $\rho$ comparable to pure WZ NWs,
demonstrating that twinning has less effect on $\rho$ than inclusion
of extended WZ or ZB segments~\cite{ThelanderNL11}.

\begin{figure}
\includegraphics [width=0.45\textwidth]{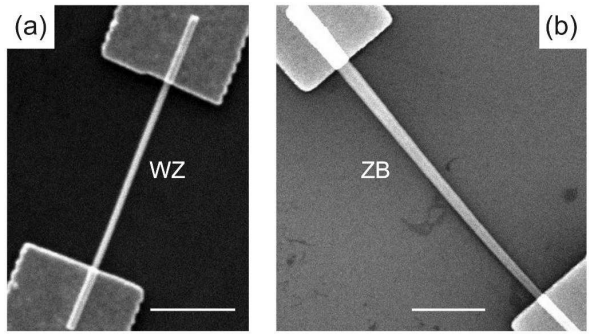}
\centering \caption{Scanning electron micrograph of completed (a)
wurtzite (WZ) and (b) zincblende (ZB) NWFETs. The slight ZB NW taper
enables confirmation of the phase species in any given NWFET.}
\label{Fig1}
\end{figure}

Here we report a study of the transfer characteristics and peak
field-effect mobility $\mu_{FE}^{pk}$ versus $T$ for NWFETs made
using nominally diameter matched pure WZ and pure ZB InAs nanowires
grown by MOVPE using the optimized conditions identified by Joyce
{\it et al.}~\cite{JoyceNL10}.

\section{Methods}

\begin{figure*}[h]
\includegraphics*[width=\textwidth]{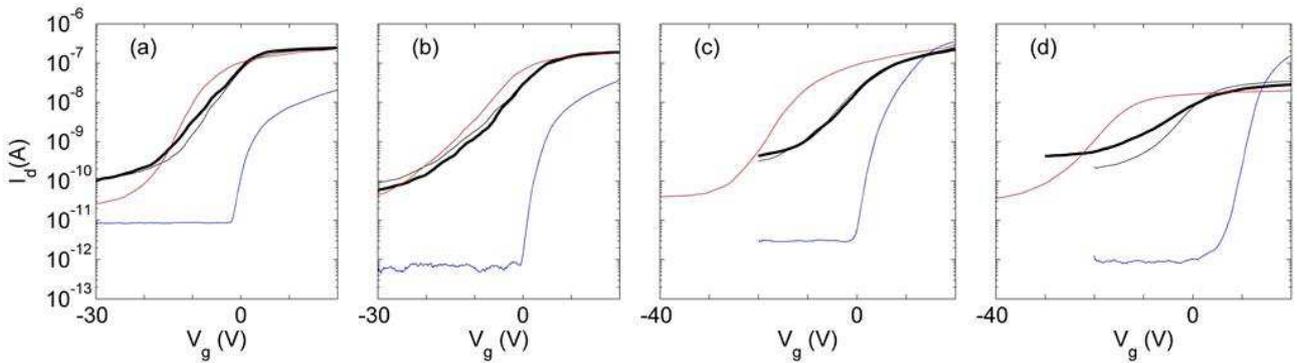}
\centering \caption{NWFET channel current $I$ versus gate voltage
$V_{g}$ at $V_{sd} = 10$~mV for devices (a) WZ-A, (b) WZ-B, (c) ZB-A
and (d) ZB-B. Data obtained in the order: $T = 300$~K/air (thin
black), $T = 75$~K/He (blue), $T = 265$~K/He (red) and $T =
300$~K/air (thick black).} \label{Fig2}
\end{figure*}

InAs NWs were grown by horizontal flow metalorganic chemical vapor
deposition (MOCVD) using trimethylindium (TMI) and AsH$_{3}$. Prior
to growth, InAs(111)B substrates were treated with poly-L-Lysine
solution followed by a solution of $50$~nm diameter Au
nanoparticles. The substrate was then annealed at $600^{\circ}$C in
AsH$_{3}$ to desorb surface contaminants. Growth was performed at
$400^{\circ}$C ($500^{\circ}$C) for ZB (WZ) with TMI flow rate $1.2
\times 10^{-5}$~mol/min and AsH$_{3}$ flow rates $5.5 \times
10^{-4}$~mol/min ($3.5 \times 10^{-5}$~mol/min). This gives V/III
ratios of 46 and 2.9 for ZB and WZ. Growth was performed for $30$
min at a pressure of $100$~mbar with H$_{2}$ carrier gas added to
give $15$~slm total flow rate. Our WZ and ZB NWs are exact replicas
of those in Figs.~1(d)/2(d) and Figs.~1(a)/2(a) of
Ref.~\cite{JoyceNL10}. These growths are reproducible, giving pure
WZ NWs free of stacking faults and pure ZB NWs without planar
crystallographic defects.

NWs were transferred onto a device substrate consisting of a
degenerately doped Si wafer capped with $100$~nm SiO$_2$/$10$~nm
HfO$_2$, and prepatterned with $5$~nm Ti/$100$~nm Au interconnects.
The doped substrate serves as the gate. Source and drain contacts
were defined by electron beam lithography (EBL). Contact passivation
was performed by immersion in (NH$_{4}$)$_{2}$S$_{x}$
solution~\cite{SuyatinNanotech07} at $40^{\circ}$C for $120$~s,
which consists of $5$~g S$_{2}$ powder added to $52$~mL $20\%$
(NH$_{4}$)$_{2}$S solution, diluted $1:200$ with deionized H$_{2}$O
immediately prior to use. $25$~nm Ni/$75$~nm Au contacts were
deposited by vacuum evaporation immediately thereafter.

Devices were made as `chips' containing two separate sets of EBL
fields. WZ NWs (ZB NWs) were deposited into the first (second) set
of EBL fields, enabling parallel fabrication of WZ and ZB NWFETs to
ensure common contact properties. The two sets were separated after
fabrication and packaged individually in $20$-pin ceramic chip
carriers. We used scanning electron microscopy (SEM) to confirm a
NWFET's phase after electrical measurements were completed; WZ and
ZB NWs can be distinguished by taper (see Fig.~1). Our key
conclusions have been corroborated in multiple NWFETs from more than
one chip.

Low $T$ electrical measurements were performed using a `dipstick' in
a liquid He dewar. We monitored $T$ using a Cernox resistor, with $T
= 4.2$~K achieved by immersion in liquid and $T
> 5$~K attained using the temperature gradient of the He
dewar atmosphere. The NWFET channel current $I$ was measured using
a.c. lock-in techniques with a source-drain voltage $V_{ds} = 10$
(\S3.1) or $4$~mV (\S3.2) at $73$~Hz. The gate voltage $V_{g}$ was
supplied by a Keithley 2400 enabling continuous gate leakage current
$I_{g}$ monitoring, with $I_{g} < 10$~nA during all measurements.

\section{Results and Discussion}
\subsection{Comparison of ZB and WZ NWFET transfer characteristics}

Figure~\ref{Fig2} shows the $I$ versus $V_{g}$ (transfer)
characteristics of two ZB and two WZ InAs NWFETs measured in air at
$T = 300$~K, and in He at $T = 265$ and $75$~K. The WZ NWFETs show
consistent characteristics; we obtain
$I_{\textrm{on}}/I_{\textrm{off}} \sim 2000$ and threshold voltage
$V_{\textrm{th}} \sim 0$~V for WZ-A, WZ-B and several other devices
on the same chip (not shown). Atmospheric composition and $T$ both
have comparatively weak effects on $V_{\textrm{th}}$. In contrast,
the ZB NWFET characteristics are more variable, with
$V_{\textrm{th}}$ changing by several volts between devices on the
same chip under common conditions, as well as with atmosphere and
$T$ for a given NWFET.

One aspect notably different from previous studies is the
subthreshold behaviour. We always obtain
$I_{\textrm{on}}/I_{\textrm{off}} > 10^{2}$ for our ZB NWFETs; the
on-off ratio for our WZ NWFETs generally tends to be slightly higher
than for our ZB NWFETs. This is in contrast to the
$I_{\textrm{on}}/I_{\textrm{off}} < 2$ for ZB InAs NWFETs in
Ref.~\cite{DayehAFM09}. Dayeh {\it et al.} attribute the poorer
subthreshold characteristics in their ZB NWFETs to electron
accumulation caused by positive surface-state charge. Our data in
Figs.~2(c/d) supports the conclusion that the ZB NWFET
characteristics are more heavily influenced by surface effects, with
greater variations observed in the off-current between air and He
atmospheres than for the WZ NWFETs (Fig.~2(a/b)). One explanation is
that ZB NWs contain micro-facets with different Miller indices,
unlike WZ NWs~\cite{JoyceNL10, SunNL12}. In addition to an increased
density of surface states, this may result in differences in In/As
surface ratio~\cite{SunNL12}, surface chemistry and surface-state
energy spectrum; surface orientation is known to affect surface
states and gate behaviour for GaAs, for example~\cite{BurkePRB12}.

Dayeh {\it et al.} also propose that the improved subthreshold
characteristics for WZ NWFETs arise from stacking faults, with
polarization charge at WZ/ZB stacking fault interfaces giving rise
to an axial `sawtooth' potential that `cuts through' the channel
accumulated by the positive surface charge to ensure complete
depletion~\cite{DayehAFM09}. One would thus expect the on-off ratio
for pure WZ nanowires (i.e., without stacking faults) to be poor. We
instead obtain high on-off ratios for our pure WZ NWFETs
(Fig.~2(a/b)), which suggests that the off-current difference
between WZ and ZB NWFETs goes beyond the spontaneous polarization
charge mechanism proposed in Ref.~\cite{DayehAFM09}. Instead, it
might simply arise from surface-state density/spectrum differences
between WZ and ZB NWs. Note well, we do not claim stacking fault
interface polarization charge provides no improvement in
subthreshold characteristics at all; heterostructure barriers
improve NWFET characteristics~\cite{FrobergIEEE-EDL08} and there is
no reason to expect a WZ/ZB  barrier would not do
likewise~\cite{DickNL10}. We only suggest that spontaneous
polarization charge is not necessary for high on-off ratio in WZ
InAs nanowires.

\subsection{Electrical Mobility versus Temperature for WZ and ZB InAs NWFETs}

\begin{figure}
\includegraphics[width=0.5\textwidth]{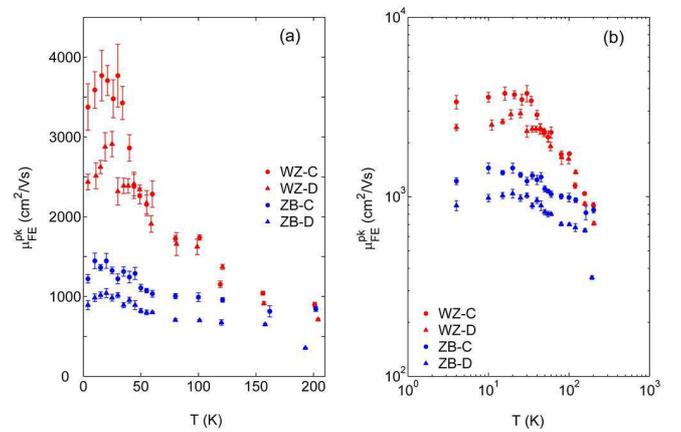}
\centering \caption{Peak field-effect mobility $\mu_{FE}^{pk}$ vs
temperature $T$ on (a) linear and (b) log-log axes for devices WZ-C
(red circles), WZ-D (red triangles), ZB-C (blue circles) and ZB-D
(blue triangles). Data obtained with $V_{sd} = 4$~mV.} \label{Fig3}
\end{figure}

To further characterize the transport differences between ZB and WZ
NWs, we obtained transfer characteristics at various $4 < T < 200$~K
for devices WZ-C, WZ-D, ZB-C and ZB-D. The field-effect mobility was
obtained as $\mu_{FE} = g_{m}L_{G}^{2}/CV_{DS}$, where $g_{m} =
\partial I_{DS}/\partial V_{GS}$ is the transconductance, $L_{G}$ is
the channel length and $V_{DS} = 4$~mV. The capacitance $C$ was
estimated using a cylinder-on-plane model~\cite{DayehSmall07}. We
obtain both $L_{G}$ and the NW diameter $d$ by SEM after electrical
measurements are completed; for the ZB NWs, we assume linear taper
and take $d$ as the average of the diameters immediately adjacent to
the source and drain contacts. We deal with the $100$~nm
SiO$_{2}$/$10$~nm HfO$_{2}$ insulator by assuming a $110$~nm layer
with dielectric constant $\epsilon_{\textrm{eff}} = 4.22$.

In Fig.~3(a/b) we plot the peak field-effect mobility
$\mu_{FE}^{pk}$ vs $T$ on linear and log-log scales for clarity. For
both WZ and ZB there is a peak in $\mu_{FE}^{pk}$ versus $T$ at $T =
20-30$~K, we attribute the low and high $T$ drop-offs in
$\mu_{FE}^{pk}$ to ionized impurity and phonon scattering, as per
bulk InAs~\cite{ConwellPR50, RodePRB71}. At low $T$, $\mu_{FE}^{pk}$
is $2-4\times$ higher for WZ than ZB; the $\mu_{FE}^{pk}$ values
converge as $T \rightarrow 200$~K. This is consistent with the
similar mobilities obtained for ZB and WZ NWs at $T =
300$~K~\cite{DayehAFM09}, and with the finding that WZ and ZB NWs
have relatively similar phonon spectra~\cite{ZhouPRB11}. The higher
low $T$ $\mu_{FE}^{pk}$ for WZ is not surprising; the lower growth
temperature for our ZB NWs will give a higher carbon background
impurity level~\cite{ThelanderNanotech10} and therefore increased
ionized impurity scattering. Surface roughness scattering may also
be partially responsible for the reduced low $T$ $\mu_{FE}^{pk}$ for
ZB NWs in Fig.~3(a/b), owing to the micro-faceting discussed
earlier~\cite{JoyceNL10}. Commenting briefly on the functional
relationships between $\mu_{FE}^{pk}$ and $T$, one expects
$\mu_{FE}^{pk} \propto T^{3/2}$ and $T^{-3/2}$ in the low and high
$T$ limits~\cite{ConwellPR50, RodePRB71}. We find $\mu_{FE}^{pk}
\propto T^{0.08}$ for both in the low $T$ limit and $T^{-0.8}$ and
$T^{-0.3}$ for WZ and ZB, respectively, in the high $T$ limit. These
differ significantly from expectations for bulk materials;
NW-specific calculations of these exponents are not available, but
would be an interesting contribution.

\section{Conclusions}

We made a comparative study of NWFETs made using pure WZ and pure ZB
InAs NWs with nominally identical diameter grown using
MOCVD~\cite{JoyceNL10} to establish how the growth differences
influence device performance. The WZ NWFETs show more consistent
subthreshold characteristics than the ZB NWFETs, and most notably,
the ZB NWFET on-off ratio and threshold voltage is more sensitive to
temperature and atmospheric composition. This points to surface
states playing a greater role in the electronic performance of ZB
NWFETs. This may be due to ZB NW surface
micro-faceting~\cite{JoyceNL10}. We find on-off ratios $\sim 100$
for the ZB NWs, $50 \times$ greater than previously
reported~\cite{DayehAFM09}. We also found that high on-off ratios
persist in WZ NWFETs without stacking faults, demonstrating that
WZ/ZB interface polarization charge~\cite{DayehAFM09} is not a
necessary condition for good subthreshold characteristics. The two
phases have similar peak field-effect mobilities in the high
temperature limit, consistent with earlier work~\cite{DayehAFM09},
but the WZ peak mobility is $2-4\times$ higher in the low $T$ limit,
likely due to a combination of higher background impurity
levels~\cite{ThelanderNanotech10} and increased surface roughness
scattering due to micro-faceting in the ZB NWs.

\begin{acknowledgement}
This work was funded by the Australian Research Council. We thank K.
Storm and L. Samuelson for assistance with device fabrication and
M.O. Williams for assistance with preliminary studies. This work was
performed in part using the NSW and ACT nodes of the Australian
National Fabrication Facility (ANFF).
\end{acknowledgement}

\end{document}